# Quantum imaging of current flow in graphene

Jean-Philippe Tetienne[1*], Nikolai Dontschuk[2], David A. Broadway[1], Alastair Stacey[1], David A. Simpson[2,3], and Lloyd C. L. Hollenberg[1,2,3†]

[1]*Centre for Quantum Computation and Communication Technology, School of Physics, University of Melbourne, Parkville, 3010, Australia*
[2]*School of Physics, University of Melbourne, Parkville, 3010, Australia*
[3]*Centre for Neural Engineering, University of Melbourne, Parkville, 3010, Australia*

*jtetienne@unimelb.edu.au
†lloydch@unimelb.edu.au

**Since its first isolation in 2004 [1], graphene has been found to host a plethora of unusual electronic transport phenomena, making it a fascinating system for fundamental studies in condensed-matter physics as well as offering tremendous opportunities for future electronic and sensing devices [2]. However, to fully realise these goals a major challenge is the ability to non-invasively image charge currents in monolayer graphene structures and devices. Typically, electronic transport in graphene has been investigated via resistivity measurements, however, such measurements are generally blind to spatial information critical to observing and studying landmark transport phenomena such as electron guiding [3] and focusing [4], topological currents [5] and viscous electron backflow [6] in real space, and in realistic imperfect devices. Here we bring quantum imaging to bear on the problem and demonstrate high-resolution imaging of current flow in graphene structures. Our method utilises an engineered array of near-surface, atomic-sized quantum sensors in diamond, to map the vector magnetic field and reconstruct the vector current density over graphene geometries of varying complexity, from mono-ribbons to junctions, with spatial resolution at the diffraction limit and a projected sensitivity to currents as small as 1 μA. The measured current maps reveal strong spatial variations corresponding to physical defects at the sub-μm scale. The demonstrated method opens up an important new avenue to investigate fundamental electronic and spin transport in graphene structures and devices, and more generally in emerging two-dimensional materials and thin film systems.**

Resistivity transport measurements have been a powerful tool for discovering electronic phenomena in condensed matter, and in particular in graphene. However, this approach averages out spatial variations, and therefore cannot distinguish the contributions of bulk processes from those of defects and edges, which are of crucial importance in the physics of microscopic devices [7, 8]. Furthermore, real-space observation of many non-classical forms of electronic transport has remained elusive. The ability to image charge currents in graphene would thus open a new era in the study of two-dimensional (2D) electronic transport. Although progress towards this goal has been recently reported, the techniques employed so far are restricted to specific graphene systems under local perturbations [9], one-dimensional imaging [10] or low spatial resolution [11]. A general, non-invasive method offering sub-µm resolution is still lacking. In this work, we fabricate an integrated quantum imaging platform where graphene devices are defined directly onto a diamond chip containing an array of near-surface, atomic-sized magnetic sensors in the form of nitrogen-vacancy (NV) centres [12-16]. Using this platform, we image the magnetic field generated by charge currents injected into graphene ribbons and junctions and reconstruct the current density distribution [17, 18], revealing current flow features associated with sub-µm defects in the graphene structures.

The principle of the experiment is illustrated in Fig. 1a. Graphene ribbons and metallic contacts are fabricated directly onto a diamond chip which hosts a layer of NV centres embedded $\approx 20$ nm below the surface. The fabrication process involves wet transfer of monolayer graphene grown by chemical vapour deposition on a copper foil, onto the diamond substrate, and subsequent patterning via electron-beam lithography and plasma etching. The graphene-diamond platform is then mounted onto a glass cover slip equipped with a microwave (MW) resonator, and placed in a wide-field microscope operating at room temperature [16]. Fig. 1b shows an optical micrograph of the final device, where only the metallic contacts are visible on the diamond due to the very weak intrinsic absorption of graphene. A zoom of an area containing a graphene ribbon (Fig. 1c) confirms that monolayer graphene on diamond provides no measurable contrast. To visualise the graphene sheet in situ, we make use of the NV photoluminescence (PL) quenching from the graphene [19]. The NV layer is illuminated with a green laser beam, and the resulting red PL is imaged with a camera (Fig. 1d). The graphene ribbon appears as a dark area due to resonance energy transfer [20]. A line cut across the ribbon (Fig. 1e) indicates a PL reduction of $\approx 30\%$. This corresponds to a mean distance between the graphene sheet and the NV layer of $\approx 20$ nm [20], in agreement with the expected NV implantation depth (see SI).

We next injected a current through one of the graphene ribbons (Fig. 2a) and used the array of NV centres to map the induced magnetic field. The measurement consists of recording the PL intensity as a function of applied MW frequency to form an optically detected magnetic resonance (ODMR) spectrum (Fig. 2b). An external magnetic field of amplitude $B_{\text{ext}} = 10$ mT was applied in order to lift the degeneracy between the four symmetry axes of NV centres, resulting in four pairs of spin resonance transitions. This allows the vector components of the magnetic field, $(B_x, B_y, B_z)$, to be extracted in the laboratory frame [14] (see SI). Figure 2c shows the vector magnetic field maps for a graphene ribbon under an injected current $I = +0.8$ mA, normalised using a current of opposite sign, $I = -0.8$ mA. The current density distribution is then obtained by inverting the Biot-Savart law in the Fourier space [17, 18]. This provides the vector components of the lineal current density, $J_x$ and $J_y$ (Fig. 2d), with a precision estimated to $\approx 5\%$ (see SI). They can be combined in a single plot (bottom panel in Fig. 2d), where the colour codes the norm of the current density vector, $|\vec{J}|$, and the arrows indicate the direction (and relative norm) of $\vec{J}$. As a consistency check, integrating the current density over the width of the ribbon gives a total current of $\approx 0.81(4)$ mA, in agreement with the applied current. Based on the signal-to-noise ratio observed in Fig. 2d, we project a sensitivity to current densities as low as $\approx 1$ A/m under similar conditions (total acquisition time of 2 hours), which corresponds to a total current of 1 µA in a 1 µm wide ribbon.

Figure 2d reveals strong irregularities in the current density along the ribbon, mainly in the form of constrictions and holes with feature sizes imaged down to $\approx 1$ µm. To investigate the origins of these effects, we mapped the current density in two graphene ribbons (Figs. 3a and 3b) and compared the maps with the corresponding PL images (Fig. 3c), as well as with scanning electron microscopy (SEM) images of the same areas (Fig. 3d), where the relatively high electron emission of the diamond makes a clear contrast with the graphene. A clear correlation is seen between the irregular features in the current density, and bright lines or dots visible in the SEM images. These bright areas, which exhibit a range of shapes with feature sizes down to $\approx 200$ nm, are attributed to cracks and localised tears in the graphene sheet, through which the current cannot flow. Such tearing is commonly found in graphene after being transferred onto a substrate using a wet process [21]. Interestingly, the tears are also visible in the PL images. They appear as bright contrast areas, with a PL intensity similar to that measured on the bare diamond surface. This indicates that no significant PL quenching occurs in these areas, consistent with an absence of graphene. Fig. 3a also reveals a non-uniform current density in tear-free areas, where the current density is maximum at the centre of the ribbon and decreases towards the edges. This indicates a non-uniform conductivity across the ribbon cross-section, possibly associated with modulations in the carrier density. We note that the size of the smallest observable features in the current maps is limited by diffraction ($\approx 500$ nm), however super-resolution techniques could be applied, with the promise of a spatial resolution down to $\approx 20$ nm [22].

Finally, as a prelude to using the system to investigate more complex geometries, we examined the current flow at the junction between a graphene ribbon and evaporated metallic contacts (Figs. 4a and 4b). The contacts are made of Ti/Au (20 nm/50 nm), with the graphene ribbon on top of the metallic leads. The current density maps are shown in Figs. 4c and 4d for two different junctions. Along with the change of direction as the current flows from the leads (along $y$) into the graphene ribbon (along $x$), we also observe constrictions in the current flow around the contact interfaces. These features are consistent with the presence of tears along the metallic edges, which likely occurred during the transfer of graphene as it had to conform to the 70 nm step of the leads. The edge-induced tears are not visible in the SEM images (Figs. 4a and 4b), illustrating the method's ability to detect and investigate defects and transport in complex geometries not accessible to conventional techniques.

The ability to map the current flow in operating graphene devices provides an avenue for hitherto inaccessible real-space investigations of conduction in the presence of impurities, grain boundaries or ripples [7, 8, 23], and of a range of non-classical transport phenomena in the ballistic regime [3-6, 10]. Another intriguing application is the study of orbital magnetism [24], which has generated significant interest and requires spatially-resolved probing methods. The NV centres used here can sense not only quasi-static magnetic fields, but also AC fields and randomly fluctuating fields (noise) by exploiting quantum decoherence [13, 25]. This has recently enabled probing of Johnson noise in metals [26], and could be used to investigate spontaneous or driven charge fluctuations in graphene. Besides electronics, diamond magnetic imaging opens exciting opportunities in graphene spintronics. In particular, it has the sensitivity required to image spin injection in graphene [27], or detect the spin Hall effect [28]. Diamond-based imaging could also be used to detect and investigate localised magnetic moments associated with defects, impurities or edges, which is a subject of active research presently [29]. Finally, we note that our approach is not limited to graphene but is applicable to multilayer graphene, other two-dimensional materials as well as thin film solid-state systems including topological insulators and 2D electronic systems in silicon [30], provided they can be transferred or fabricated onto diamond, or onto a thin buffer layer of a different material if required. As such, the quantum imaging technique reported here could become a ubiquitous investigation tool for 2D technologies in the coming years.

## Acknowledgements

The authors would like to thank S. Rubanov from the Advanced Microscopy Facility of the Bio21 Institute, University of Melbourne, for assistance with SEM imaging. The authors also acknowledge helpful discussions with C. Ritchie and L. T. Hall. This research was supported in part by the Australian Research Council Centre of Excellence for Quantum Computation and Communication Technology (Project number CE110001027). L.C.L.H acknowledges the support of the Australian Research Council Laureate Fellowship Scheme (FL130100119). D.A.S acknowledges support from the Melbourne Neuroscience Institute Fellowship Scheme.

## Author Contributions

The experiment was designed by all authors. J.-P.T. performed the NV measurements, the numerical reconstruction, and the SEM imaging. J.-P.T. and L.C.L.H. wrote the manuscript. N.D. and A.S. fabricated the graphene devices on diamond and contributed to data interpretation. N.D., D.A.B, A.S and D.A.S prepared and characterised the diamond sample for NV imaging. D.A.S realised the imaging setup and contributed to the NV measurements. All authors discussed the data and commented on the manuscript.

## Author Information

The authors declare no competing financial interests. Correspondence and requests for materials should be addressed to J.-P.T. (jtetienne@unimelb.edu.au) or L.C.L.H. (lloydch@unimelb.edu.au).

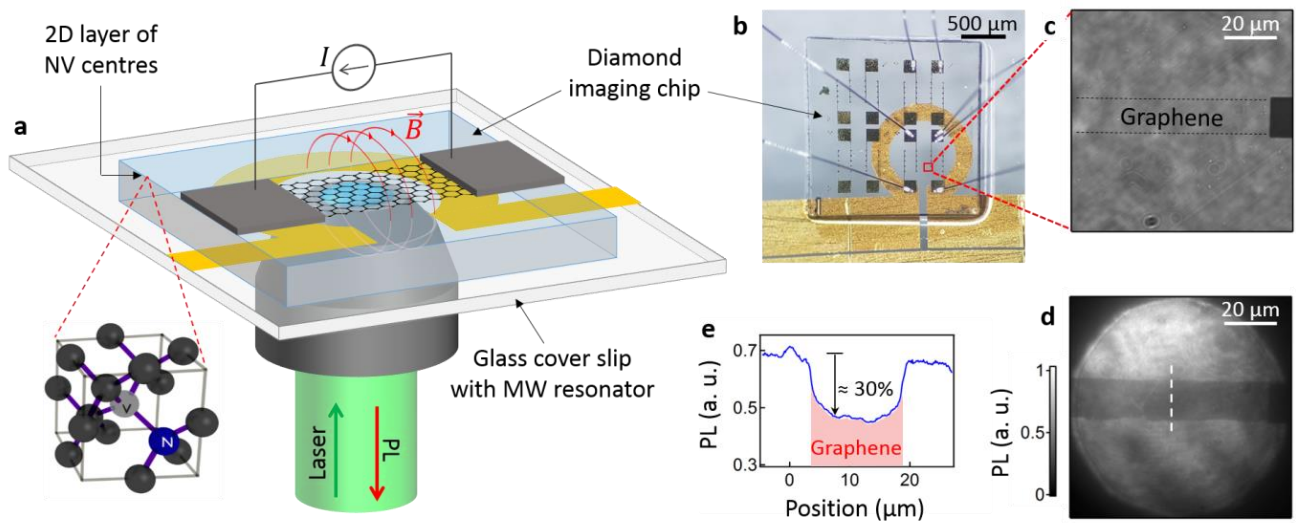

**Fig. 1 – Graphene ribbons on a diamond imaging platform.** (a) Schematic of the experiment. The diamond platform consists of a diamond chip hosting a layer of near-surface nitrogen-vacancy (NV) centres. The graphene devices are fabricated directly on the diamond chip, which is mounted on a cover slip equipped with a microwave (MW) resonator. The NV centres' photoluminescence (PL) under green laser and microwave excitations is imaged on a camera to form the magnetic field image. (b) Optical micrograph of the final device. Apparent on the diamond are metallic contacts, and wire bonds, used for current injection in the graphene ribbons. (c) Bright-field image recorded with the camera, zoomed on a graphene ribbon (not visible). (d) PL image of the same area under laser excitation. The graphene ribbon is now visible due to PL quenching. (e) Line cut across the ribbon extracted from (d) (white dashed line).

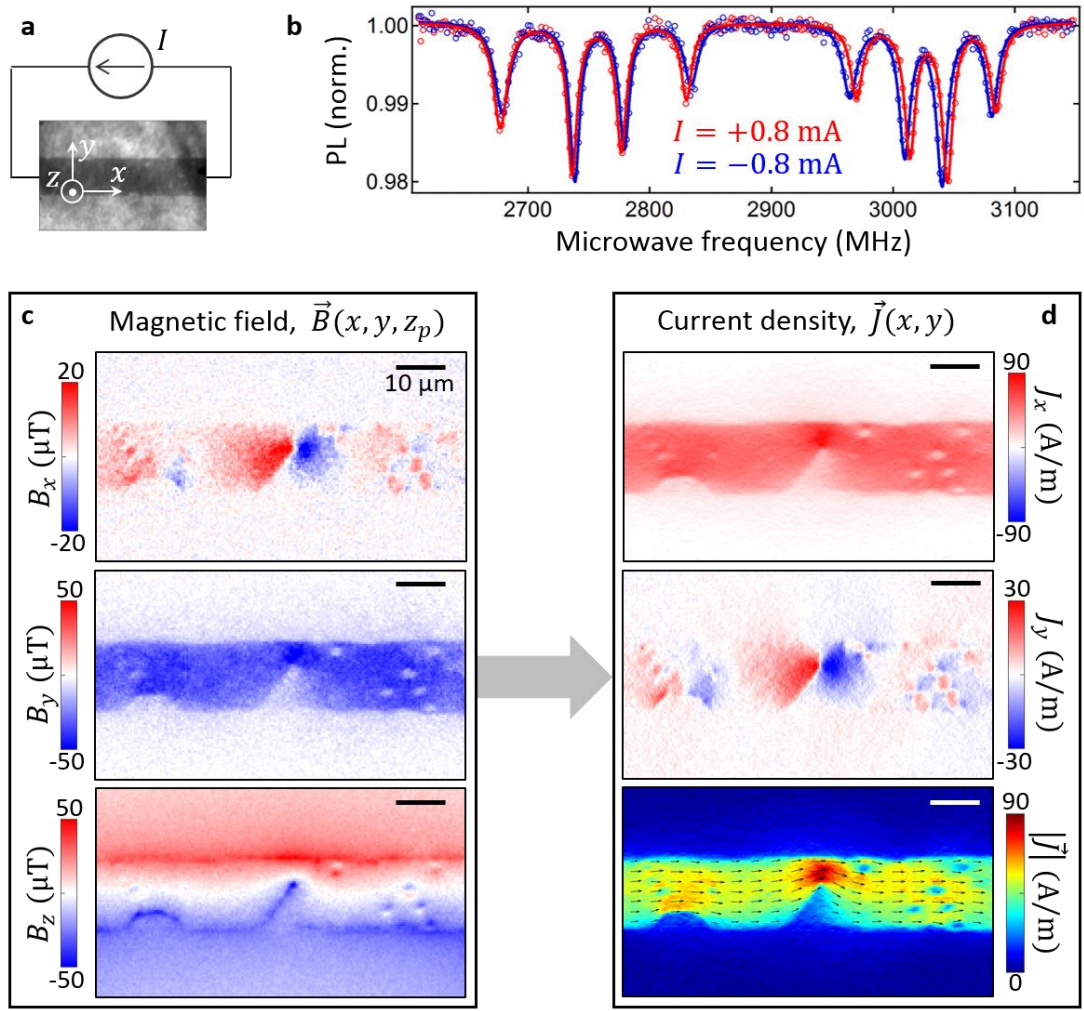

**Fig. 2 – Magnetic field imaging and reconstruction of the current density.** (a) PL image of the graphene ribbon under study, defining the $xyz$ reference frame. (b) Optically detected magnetic resonance spectrum of the NV centres in a single pixel near the graphene under a positive (red dots) or negative (blue dots) applied current. Solid lines are data fit to a sum of 8 Lorentzian functions. (c) Maps of the $B_x$ (top), $B_y$ (middle) and $B_z$ (bottom) components of the magnetic field produced by a total current $I = 0.8$ mA. (d) Maps of the $J_x$ (top) and $J_y$ (middle) components of the current density reconstructed from (c). The bottom panel shows the norm of the current density, $|\vec{J}|$. The black arrows represent the vector $\vec{J}$ (length proportional to $|\vec{J}|$, threshold $|\vec{J}| > 30$ A/m). In (c,d), all scale bars are 10 µm.

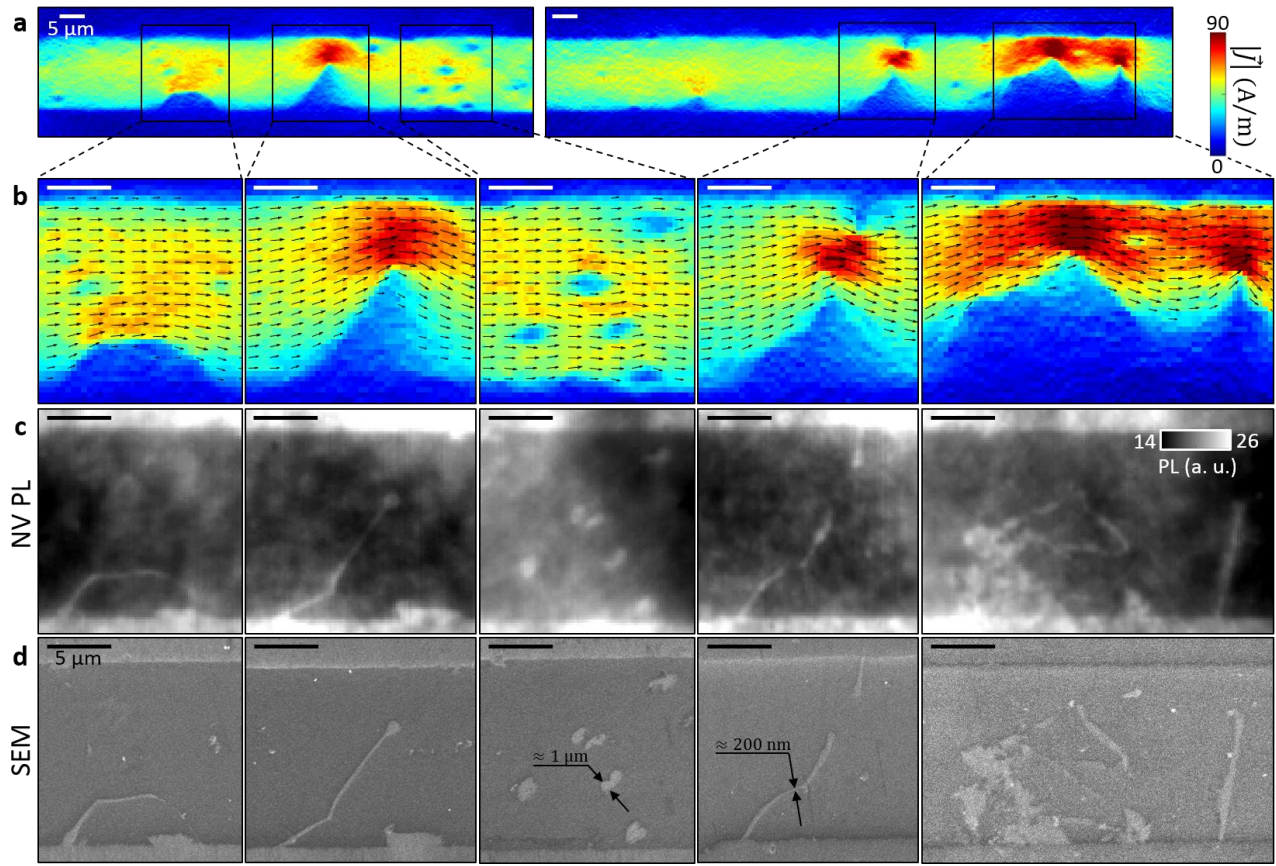

**Fig. 3 – Current flow near defects in graphene.** (a) Maps of the norm of the current density, $|\vec{J}|$, in two different graphene ribbons driven by a total current $I = 0.8$ mA. (b) Zooms of selected areas from (a). The black arrows represent the vector $\vec{J}$ (length proportional to $|\vec{J}|$, threshold $|\vec{J}| > 30$ A/m). (b) PL images corresponding to the same areas as in (b). (c) Corresponding SEM images. All scale bars are 5 µm.

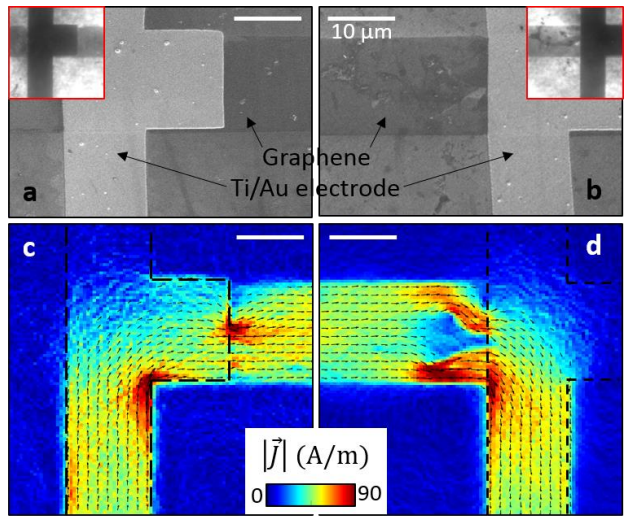

**Fig. 4 – Current flow near metallic contacts.** (a,b) SEM images of two junctions between Ti/Au electrodes and a graphene ribbon. The insets show the corresponding PL images. (c,d) Maps of the norm of the current density, $|\vec{J}|$, under a total current $I = 0.8$ mA, corresponding to the two junctions shown in (a,b). The black dashed lines indicate the edges of the metallic electrodes, as extracted from the PL images. The black arrows represent the vector $\vec{J}$ (length proportional to $|\vec{J}|$, threshold $|\vec{J}| > 30$ A/m). All scale bars are 10 µm.

# Supplementary Information for the manuscript "Quantum imaging of current flow in graphene"

Jean-Philippe Tetienne[1,*], Nikolai Dontschuk[2], David A. Broadway[1], Alastair Stacey[1], David A. Simpson[2,3], and Lloyd C. L. Hollenberg[1,2,3,†]

[1]*Centre for Quantum Computation and Communication Technology, School of Physics, University of Melbourne, Parkville, 3010, Australia*
[2]*School of Physics, University of Melbourne, Parkville, 3010, Australia*
[3]*Centre for Neural Engineering, University of Melbourne, Parkville, 3010, Australia*
[*]*jtetienne@unimelb.edu.au*
[†]*lloydch@unimelb.edu.au*

1. **Sample fabrication**

The imaging sensor used in this work is engineered from electronic grade Type IIa diamond (E6). The diamonds were thinned and repolished to a 4 x 4 x 0.1 mm crystal (DDK, USA) and then laser cut into 2x2 mm imaging chips. The single crystal diamond was then implanted with $^{15}$N atoms at 6 keV to a dose of $1 \times 10^{13}$ ions/cm$^2$. The implanted sample was annealed at 1000 °C for three hours and acid treated (sulphuric acid (1ml), sodium nitrate (1g) at 300 °C for 10 minutes) to remove any unwanted surface contamination. The density of NV centres post annealing was $1 \times 10^{11}$ NV/cm$^2$, estimated by comparing the intensity from a single NV centre in diamond with that obtained from the NV ensemble. The depth distribution of the NV centres with a 6 keV implantation energy is predicted to be $\approx 10 \pm 4$ nm by transport and range of ions in matter (TRIM) simulations. However it has been shown by molecular dynamics simulations [1], and by proton nuclear magnetic resonance measurements [2], that the NV depth can be twice as large as the TRIM predictions, leading to an estimated average depth of $\approx 20$ nm in our sample.

The graphene devices were fabricated on the diamond substrate with a two-step electron-beam lithography (EBL) process. In the first step Ti/Au 20/50nm contacts were evaporated onto the diamond surface through a soft mask patterned in a bi-layer MMA(E11)/PMMA(A4 950) resist stack. After lift-off of the resist stack in acetone, PMMA coated CVD graphene, purchased as grown on copper foil (Graphene Supermarket), was transferred using the wet chemical etch method of Liang et al. [3] without the SC1 step. After drying over 48 hours at room temperature, a layer of negative tone resist, SU8 (2002), was spun on top of the transfer PMMA(A4 950) layer. The SU8 was patterned by EBL to protect the areas of graphene to remain in the final device, whilst the exposed PMMA/graphene stack was etched away by oxygen plasma asher. Finally, the cured SU8 protective mask was removed by acetone dissolution of the underlying PMMA layer exposing the patterned graphene.

The graphene-diamond chip was then glued onto a glass coverslip equipped with a gold microwave resonator fabricated by photo-lithography. To estimate the quality of the graphene in the final device, Raman spectra under 532 nm wavelength excitation were measured at various locations. To block the NV photoluminescence, the spectra were measured from the graphene sitting on the gold contacts. Fig. S1 shows two example spectra recorded at two different spots. The bottom spectrum shows the effect of the tearing of graphene with the presence of the D peak at 1350 cm$^{-1}$. This is an disallowed transition in sp$^2$ carbon, its presence indicates edges (or other defects) that allow non-radiative conservation of angular momentum [4]. The ratio between the G (1580 cm$^{-1}$) and 2D (2600cm$^{-1}$) peaks is around 2, which is low but not uncommon for CVD single-layer graphene. The positions of the G and 2D peaks suggest relatively neutral graphene [5], under little to no strain from the substrate as expected for graphene on gold [6].

## 2. Scanning electron microscopy

The scanning electron microscopy (SEM) images in main text Figs. 3 and 4 were taken using a FEI Nova Nanolab system, using the secondary electrons in-lens detector at an acceleration voltage of 5 kV. Significant charging of the insulating diamond was observed, limiting the spatial resolution of the images. We note, however, that charging was essential in order to see the graphene. Indeed, images taken with the graphene and gold electrodes connected to the ground provided higher resolution images of the diamond surface (due to reduced charging), but exhibited no measurable contrast between graphene and bare diamond (see comparison in Fig. S2). The images shown in main text Figs. 3 and 4 were therefore taken with the sample completely isolated from the ground.

## 3. Magnetic field imaging

**Experimental details.** The wide-field magnetic imaging was performed using the setup described in Ref. [7], which is based on a modified Nikon inverted microscope (Ti-U). Optical excitation from a 532 nm Verdi laser was focused ($f$ = 300 mm) onto an acousto-optic modulator (Crystal Technologies Model 3520-220) and then expanded and collimated (Thorlabs beam expander GBE05-A) to a beam diameter of 10 mm. The collimated beam was focused using a wide-field lens ($f$ = 300mm) to the back aperture of the Nikon x40 (1.3 NA) oil immersion objective via a Semrock dichroic mirror (Di02-R561-25x36). The NV fluorescence was filtered using two bandpass filters before being imaged using a tube lens ($f$ = 300mm) onto a sCMOS camera (Neo, Andor). Microwave excitation was provided by an Agilent microwave generator (N5182A) and switched using a Minicirutis RF switch (ZASWA-2-50DR+). The microwaves were amplified (Amplifier Research 20S1G4) before being sent to the microwave resonator. A Spincore PulseBlaster (ESR-PRO 500MHz) was used to control the timing sequences of the excitation laser, microwaves and sCMOS camera and the images were obtained and analysed using a combination of custom LabVIEW/Matlab codes. The laser power density used for imaging was 30 W/mm$^2$ and all images were taken in an ambient environment at room temperature. The graphene devices were connected to a Picoammeter/Voltage Source (Keithley 6487) to operate the DC current through the devices. The magnetic images were obtained by recording optically detected magnetic resonance (ODMR) spectra of the NV layer in the pulsed regime [8], with a $\pi$-time of 200 ns. The acquisition time was 24 seconds per microwave frequency, hence 2 hours in total for the whole spectrum. A background magnetic field from a permanent magnet (strength 10 mT at the sample) was applied in such a way that all 4 pairs of spin resonances, corresponding to the four possible NV centre's symmetry axes, could be resolved (see an example spectrum in main Fig. 2b).

**Analysis of the ODMR spectra.** The ODMR spectrum for each pixel was fit to a sum of eight Lorentzian functions to obtain the resonance frequencies, $\{\omega_{i\pm}\}_{i=1 \text{ to } 4}$, with a $1\sigma$ uncertainty on each frequency of $\approx 0.1$ MHz. In general, $\omega_{i\pm}$ are non-trivial functions of $B_{\parallel,i}$ and $B_{\perp,i}$, defined as the parallel and perpendicular components of the local magnetic field with respect to the $i$th NV axis, respectively [9, 10]. They can be obtained from the spin Hamiltonian of the NV centre's ground state [11],

$$\frac{\mathcal{H}_i}{\hbar} = DS_{Z,i}^2 + \gamma_e \left( B_{\parallel,i} S_{Z,i} + B_{\perp,i} S_{X,i} \right), \tag{1}$$

where $D = 2\pi \times 2870$ MHz is the NV zero-field splitting, $\vec{S}_i = (S_{X,i}, S_{Y,i}, S_{Z,i})$ is the spin-1 operator with $Z, i$ referring to the $i$th NV axis, and $\gamma_e$ is the electron's gyromagnetic ratio. The resonance frequencies, $\omega_{i\pm}$, correspond to the differences between the eigenvalues of $\mathcal{H}_i$. Due to the first term in Eq. (1), $\omega_{i\pm}$ are mostly sensitive to $B_{\parallel,i}$ while varying relatively slowly with $B_{\perp,i}$ in the range of fields 0-10 mT. Using the exact expressions for $\omega_{i\pm}$ to extract $B_{\parallel,i}$ and $B_{\perp,i}$ would then lead to large uncertainties given the small shifts ($< 1$ MHz) induced by the Oersted field and the experimental uncertainty on $\omega_{i\pm}$. To circumvent this problem, we reduce the number of parameters by considering the difference $\Delta\omega_i = \omega_{i+} - \omega_{i-}$ (the Zeeman splitting), which can be expressed to a good approximation as a function of $B_{\parallel,i}$ alone, via the relation

$$\Delta\omega_i = \omega_{i+} - \omega_{i-} \approx 2\gamma_e B_{\parallel,i}. \tag{2}$$

Fig. S3a shows $\Delta\omega_i$ as a function of to $B_{\parallel,i}$ and $B_{\perp,i}$, as computed exactly from Eq. (1), illustrating that $\Delta\omega_i$ is mainly a function of $B_{\parallel,i}$ with little variation with $B_{\perp,i}$. Fig. S3b shows that Eq. (2) is an excellent approximation for $B_{\parallel,i} > 2$ mT. This allows us to extract $B_{\parallel,i}$, that is, the projection of the local magnetic field along the $i$th NV axis, using the approximate expression

$$B_{\parallel,i} \approx \frac{\omega_{i+} - \omega_{i-}}{2\gamma_e}. \tag{3}$$

Under the conditions of main Figs. 2-4, knowing the direction and amplitude of the applied magnetic field, we find that the error introduced by the use of this approximation in estimating the projections $\{B_{\parallel,i}\}$ is $< 1\%$ for 3 of the 4 NV axes, and $\approx 2\%$ on the 4$^{\text{th}}$ axis. In comparison, the uncertainty on $\omega_{i\pm}$ from fitting the ODMR spectrum leads to an uncertainty of $\approx 5$ μT on $B_{\parallel,i}$, which corresponds to a relative uncertainty of 10% for the maximum values measured, e.g. in Fig. S5a. We therefore use Eq. (3) to deduce the four projections $\{B_{\parallel,i}\}_{i=1 \text{ to } 4}$, which will be denoted $\{B_{\text{NV},i}\}$ in what follows.

The measured projections $\{B_{\text{NV},i}\}$ comprise the Oersted magnetic field, which depends on the applied current, $I$, and the external magnetic field, which forms a non-uniform background independent of the value of $I$ (Fig. S4, leftmost panel). To suppress this background, we record images with two opposite signs of $I$, and use the normalised difference

$$\Delta B_{\text{NV},i}(+I) = \frac{B_{\text{NV},i}(+I) - B_{\text{NV},i}(-I)}{2}. \tag{4}$$

The image $\Delta B_{\text{NV},i}(+I)$ thus represents the magnetic field projection due to the Oersted field produced by a current $+I$, free of any background (Fig. S4, rightmost panel). Applying this normalisation to the four NV axes provides the projections $\{\Delta B_{\text{NV},i}\}_{i=1 \text{ to } 4}$ of the Oersted magnetic field, as illustrated in Fig. S5a.

**Reconstruction of the vector field.** The next step is to deduce the Cartesian components of the Oersted magnetic field in the lab frame, $\{B_x, B_y, B_z\}$, where $z$ is the normal to the diamond surface and $x$ is parallel to the edges of the graphene ribbons (see Fig. S4a). The diamond chip has a (100) top surface, and the graphene ribbons are patterned relative to the diamond crystal such that the Cartesian coordinates of the NV axes have direction vectors $\{\vec{u}_{\text{NV},i}\}$ given by

$$\vec{u}_{\text{NV},1} = \frac{1}{\sqrt{2}}(0,1,1)$$
$$\vec{u}_{\text{NV},2} = \frac{1}{\sqrt{2}}(0,-1,1)$$
$$\vec{u}_{\text{NV},3} = \frac{1}{\sqrt{2}}(1,0,1)$$
$$\vec{u}_{\text{NV},4} = \frac{1}{\sqrt{2}}(-1,0,1). \tag{5}$$

Each $\Delta B_{\text{NV},i}$ image is matched to one of the above direction vectors by simply examining the magnetic field distribution. The scalar projections $\{\Delta B_{\text{NV},i}\}_{i=1 \text{ to } 4}$ and the vector components of $\vec{B} = (B_x, B_y, B_z)$ are related by a set of four equations,

$$\Delta B_{\text{NV},i} = \vec{B} \cdot \vec{u}_{\text{NV},i} \text{ for } i = 1 \text{ to } 4. \tag{6}$$

We use a least-square minimisation algorithm to obtain a best estimate of $\{B_x, B_y, B_z\}$ for each pixel, which gives an uncertainty on each component of $\approx 2$ μT. The resulting images are shown in Fig. S5b for the data of Fig. S5a.

### 4. Reconstruction of the current density

**General method.** The Oersted field $\vec{B}(\vec{r})$ originates from a current characterised by the current density $\vec{J}(\vec{r})$. Assuming a quasistatic current, $\vec{B}$ and $\vec{J}$ are related by the Biot-Savart law,

$$\vec{B}(\vec{r}) = \frac{\mu_0}{4\pi} \int \frac{\vec{J}(\vec{r}') \times (\vec{r} - \vec{r}')}{|\vec{r} - \vec{r}'|^3} d^3\vec{r}', \tag{7}$$

where $\mu_0$ is the permeability of free space. Here the current is confined in the 2D sheet of graphene, which lies in the $xy$ plane ($z = 0$). Therefore, $\vec{J}$ can be redefined as a lineal current density expressed in units of A/m, which depends only on the planar coordinates $(x, y)$, the integral in Eq. (7) becoming a 2D integral over $(x', y')$. We can therefore express the magnetic field components as

$$B_x(x,y,z) = \frac{\mu_0 z}{4\pi} \int_{-\infty}^{+\infty} \int_{-\infty}^{+\infty} \frac{J_y(x',y')}{[(x-x')^2 + (y-y')^2 + z^2]^{3/2}} dx' dy'$$

$$B_y(x,y,z) = \frac{\mu_0 z}{4\pi} \int_{-\infty}^{+\infty} \int_{-\infty}^{+\infty} \frac{-J_x(x',y')}{[(x-x')^2 + (y-y')^2 + z^2]^{3/2}} dx' dy' \quad (8)$$

$$B_z(x,y,z) = \frac{\mu_0}{4\pi} \int_{-\infty}^{+\infty} \int_{-\infty}^{+\infty} \frac{J_x(x',y')(y-y') - J_y(x',y')(x-x')}{[(x-x')^2 + (y-y')^2 + z^2]^{3/2}} dx' dy' .$$

The problem we seek to solve is to reconstruct the 2D current density $\vec{J}(x,y)$ using a 2D measurement of the magnetic field at a given distance $z_p$ above the plane of the current flow, i.e. $\vec{B}(x,y,z=z_p)$. A simple way to tackle this problem is given in Ref. [12], which relies on expressing Eqs. (8) in the reciprocal space of the $xy$ plane. We define the 2D Fourier transforms as

$$b_k(k_x, k_y, z_p) = \int_{-\infty}^{+\infty} \int_{-\infty}^{+\infty} B_k(x, y, z_p) e^{i(k_x x + k_y y)} dx\, dy \quad \text{for } k = x, y, z \quad (9)$$

$$j_k(k_x, k_y) = \int_{-\infty}^{+\infty} \int_{-\infty}^{+\infty} J_k(x, y) e^{i(k_x x + k_y y)} dx\, dy \quad \text{for } k = x, y, \quad (10)$$

where $k_x$ and $k_y$ are the components of the spatial frequency $\vec{k}$. Using these new variables, Eqs. (8) become [12]

$$b_x(k_x, k_y, z_p) = \frac{\mu_0}{2} e^{-z_p\sqrt{k_x^2+k_y^2}} j_y(k_x, k_y) \quad (11)$$

$$b_y(k_x, k_y, z_p) = -\frac{\mu_0}{2} e^{-z_p\sqrt{k_x^2+k_y^2}} j_x(k_x, k_y) \quad (12)$$

$$b_z(k_x, k_y, z_p) = i\frac{\mu_0}{2} e^{-z_p\sqrt{k_x^2+k_y^2}} \left[ \frac{k_y}{\sqrt{k_x^2+k_y^2}} j_x(k_x, k_y) - \frac{k_x}{\sqrt{k_x^2+k_y^2}} j_y(k_x, k_y) \right]. \quad (13)$$

In addition, under the quasistatic assumption, the current density obeys the equation of continuity, $\vec{\nabla} \cdot \vec{J} = 0$, which in the reciprocal space becomes

$$k_x j_x(k_x, k_y) + k_y j_y(k_x, k_y) = 0. \quad (14)$$

Thus, for each spatial frequency $(k_x, k_y)$, Eqs. (11-14) form a set of four equations for two unknowns, $j_x(k_x, k_y)$ and $j_y(k_x, k_y)$. Since this is an overdetermined system, one can choose to use only any two of the four equations, or to use redundant equations together with a least-square minimisation algorithm to obtain a best estimate of the unknowns. Once the functions $j_x(k_x, k_y)$ and $j_y(k_x, k_y)$ have been determined, we use the inverse Fourier transform to obtain the current density components in the real space,

$$J_k(x, y) = \frac{1}{(2\pi)^2} \int_{-\infty}^{+\infty} \int_{-\infty}^{+\infty} j_k(k_x, k_y) e^{-i(k_x x + k_y y)} dk_x dk_y \quad \text{for } k = x, y, \quad (15)$$

and deduce the norm of the current density vector, $|\vec{J}(x,y)| = \sqrt{J_x(x,y)^2 + J_y(x,y)^2}$.

**Numerical reconstruction.** In practice, we used the program Matlab to apply the reconstruction procedure to the experimental data. The various steps are illustrated in Fig. S6 on one particular example. Starting from the measured maps of $B_x$, $B_y$ and $B_z$ (Fig. S6a), we first compute the Fourier transforms $b_x$, $b_y$ and $b_z$ using a fast Fourier transform algorithm (Fig. S6b). Next, for each spatial frequency $(k_x, k_y)$, we apply a least-square minimisation algorithm to the system of overdetermined Eqs. (11-14), and obtain a best estimate of $j_x$ and $j_y$ (Fig. S6c). An inverse fast Fourier transform is applied to $j_x$ and $j_y$ to obtain $J_x$ and $J_y$ (Fig. S6d), which can also be represented as an image of the norm $|\vec{J}|$ with arrows to indicate the vector direction (Fig. S6e). As a consistency check, we apply Eqs. (11-13) to $j_x$ and $j_y$ to obtain the corresponding (reconstructed) magnetic field in the Fourier space (Fig. S6f) and then in the real space (Fig. S6g), which is found to match the original data (Fig. S6a) within the experimental uncertainty (2 µT). The overall procedure takes about a minute to compute on a standard computer.

**Robustness against choice of equations.** We now analyse how the reconstructed current density is affected by the choice of equations, among the four available from Eqs. (11-14). To this aim, we compare the maps obtained on one particular example under different combinations of equations (Fig. S7a), with Fig. S7b showing line cuts across the ribbon. Since there are two unknowns, at least two equations must be used. Recalling, Eq. (11) utilises the $b_x$ map, Eq. (12) utilises the $b_y$ map, Eq. (13) utilises the $b_z$ map, and Eq. (14) is the equation of continuity for the current density, $\vec{\nabla} \cdot \vec{J} = 0$. Overall, the various maps look similar under any combination of equations, differing only in the noise level and in small deviations (Fig. S7a). Using only $b_x$ and $b_y$ gives the largest noise, however adding the continuity equation helps to dramatically reduce the noise, but also reduces the values of current density in the ribbon by $\approx 6\%$ (see Fig. S7b, green curve vs. black curve). On the contrary, adding the $b_z$ equation to $b_x$ and $b_y$, without current continuity, gives larger values by $\approx 3\%$ (blue curve). On the other hand, using only $b_z$ together with the continuity equation produces an artefact in that the current density increases towards the edge of the images (far from the ribbon) where it should be null (pink curve in Fig. S7b). We attribute this effect to long range contributions to the out-of-plane ($B_z$) field component caused by current flow in the leads. These contributions affect mainly $B_z$ which decays slowly over long distance (as $1/r$), as opposed to $B_x$ and $B_y$ which vanish outside the current-carrying structures (see Eq. (16) and Fig. S8). In order to mitigate the long range contributions in the $b_z$ equation, we added a weight to this equation so that it is effectively utilised only for spatial frequencies above a certain threshold. This solution, which was used to produce all current density maps in the main paper, allows us to adequately benefit from all information available, i.e. to use the four equations, wherever suitable in the frequency space. The threshold for the weight was chosen so as to produce the sharpest decay of the current density outside the ribbon (cyan curve in Fig. S7b).

Although the reconstruction is relatively robust, the fact that different combinations of equations produce variations in the reconstructed current density values (inside the ribbon), and that the current density does not completely vanish within a few µm outside the ribbon, suggests that the assumed model is not exactly complete, i.e. the measured magnetic field comprises small additional contributions other than the Oersted field from the current flow in the graphene ribbon. Regardless of its origin (discussed below), the variability in the reconstructed current density maps enables us to estimate the precision of the reconstruction. Based on the variability in the current density inside the ribbon as seen in Fig. S7b ($\approx 54 \pm 3$ A/m), we thus attribute a relative precision of 5% to the values of current density reported in the main paper. We note that integrating the current density over the width of the ribbon gives a total current that indeed matches the applied current within 5%. As per the origin of the deviations from the Oersted model, along with long range contributions from the leads other potential causes include: (i) a magnetic response from the diamond surface, which may screen some of the components of the Oersted field, (ii) the spin Hall effect in graphene [13], which may add an additional contribution to the measured magnetic field. While these spurious contributions limit the precision in the present work, diamond-based magnetic imaging offers a unique opportunity to investigate the underlying phenomena (in particular, the spin Hall effect) in the real space.

**Robustness against choice of probe distance.** The probe distance, $z_p$, enters as a parameter in Eqs. (11-13) used for the reconstruction. In our experiments, $z_p$ corresponds to the NV-graphene distance, which takes a distribution of values due to the stochastic nature of the implantation process. As a nominal value, we used $z_p = 20$ nm in all reconstructions performed in the main paper. Here we show that the exact choice of this value has a negligible impact on the results. This is tested by performing a reconstruction using different values of $z_p$. Fig. S7c shows line cuts across the ribbon obtained with two extreme values, $z_p = 10$ and $80$ nm. The relative difference between the two curves is $\approx 1\%$, much smaller than the variations discussed previously. This justifies the use of a single value of $z_p$ to model a distribution of distances, e.g. $20 \pm 10$ nm as estimated for the graphene-NV distance in our sample. Another consequence of this insensitivity is that the same nominal value ($z_p = 20$ nm) can be used to reconstruct the 2D current density in the metallic leads, which are 70 nm thick, since the possible variation of the current density in the $z$ direction is effectively irrelevant here. This enabled us to reconstruct the current density at the lead-graphene junction in main Fig. 4, via a single reconstruction run assuming $z_p = 20$ nm.

The insensitivity of the reconstruction to the probe distance can be understood by examining analytic expressions for the Oersted field in an ideal case. Let us consider a graphene ribbon of width $W$, whose edges are parallel to $x$. We assume a current of total intensity $I$ flowing along the $x$ direction, with a uniform density $J = I/W$ across the width of the ribbon (i.e., in the $y$ direction). Integrating Eqs. (8) over the ribbon gives the field components

$$B_x(y, z_p) = 0$$

$$B_y(y, z_p) = -\frac{\mu_0 J}{2\pi}\left[\tan^{-1}\left(\frac{W-2y}{2z_p}\right) + \tan^{-1}\left(\frac{W+2y}{2z_p}\right)\right] \qquad (16)$$

$$B_z(y, z_p) = \frac{\mu_0 J}{4\pi}\ln\left[\frac{(W+2y)^2 + 4z_p^2}{(W-2y)^2 + 4z_p^2}\right],$$

where the ribbon is centred at $y = 0$. Fig. S8 show profiles of those components across the ribbon, for various probe distances, $z_p = 20, 50, 80$ nm. The difference between the different curves is limited to the very edges of the ribbon, as shown in the insets. However, the reconstruction of the current density relies on the full measured magnetic field maps, not just at the edges of the ribbon. Moreover, the deviations occur where the field exhibits sharp variations of the magnetic field, over length scales of $\approx 500$ nm. This length scale is of the order of the pixel size in the measured magnetic field images (430 nm × 430 nm), which also approximately matches the spatial resolution (limited by diffraction) of the optical microscope used to readout the magnetic field. As a consequence, since the maximum spatial frequency used in the reconstruction of the current density is given by this inverse of the pixel size, higher spatial frequencies will not contribute to the reconstruction, and therefore the fine differences observed in Fig. S8 for different probe distances are effectively irrelevant. This explains why the choice of the probe distance (in the range 10-100 nm) has a negligible impact on the reconstructed current density maps, as illustrated in Fig. S7c.

**Robustness against other parameters.** We verified that an error on the pixel size of the real-space images has a negligible impact on the reconstructed current density maps: a 20% error on the pixel size changes the current density values by $< 1\%$. We also tested the effect of the finite size of the real-space images in computing the Fourier transforms. To do so, we expand the size of the $B_x$, $B_y$ and $B_z$ matrices and pad them with zeros before applying the fast Fourier transform algorithm. We find that increasing the size by a factor up to 4 (in each spatial direction) changes the current density values by $< 2\%$, with good convergence for a factor 2 and above. In all reconstructions performed in the main paper, we used a factor 2 as this gives the best compromise between computation speed and precision.

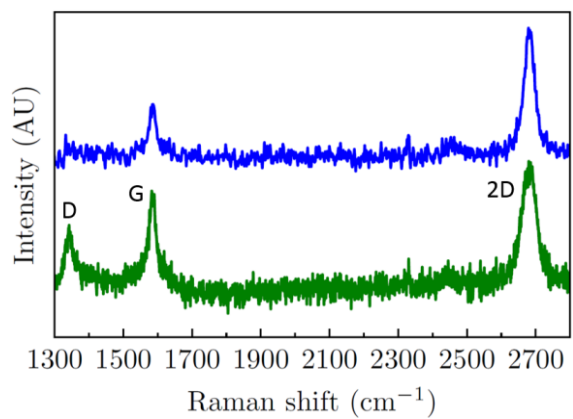

**Figure S1 – Raman spectroscopy.** The two spectra correspond to two different spots on the graphene on the gold electrodes. The excitation light has a 532 nm wavelength. The main peaks are labelled.

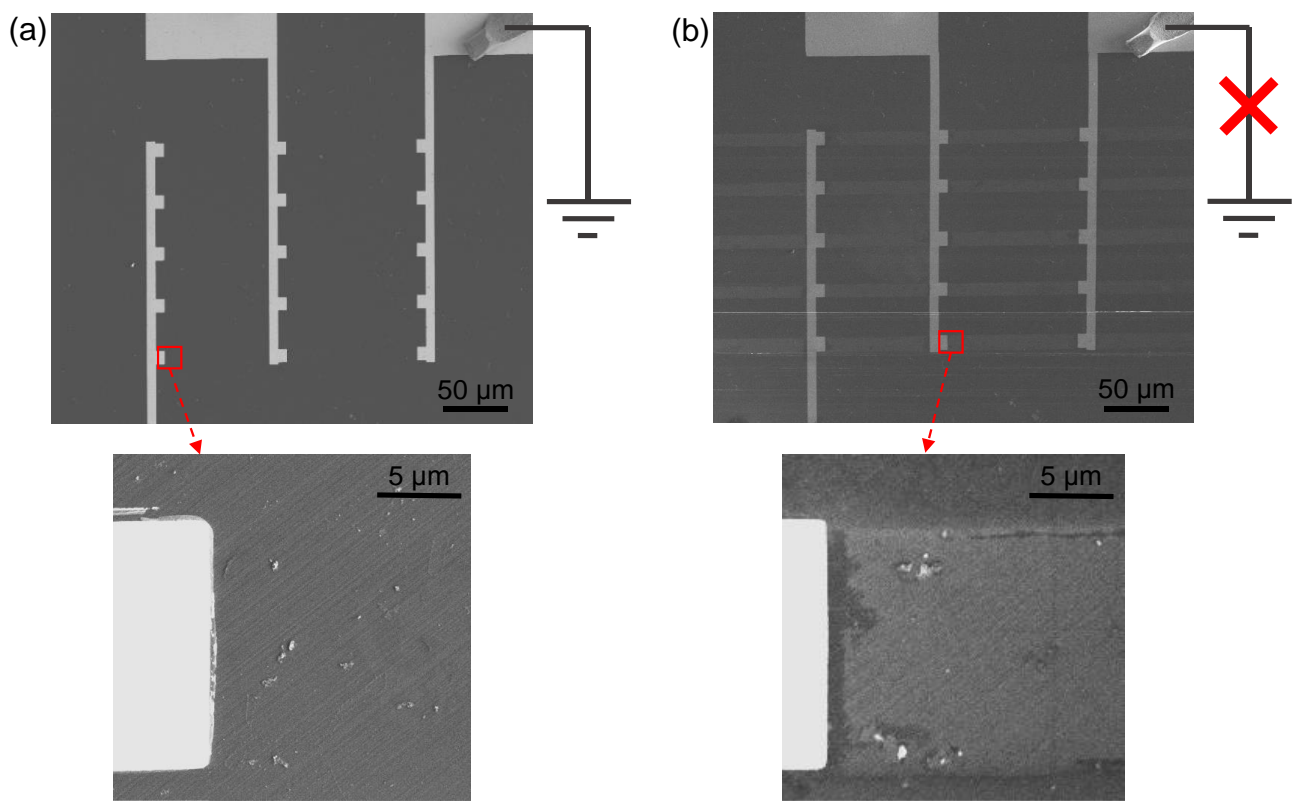

**Figure S2 – SEM imaging of graphene on diamond.** (a) SEM images taken with the electrodes connected to the ground via the metallic sample holder. (b) SEM images taken with the electrodes electrically isolated from the ground. Note that the diamond substrate is insulating.

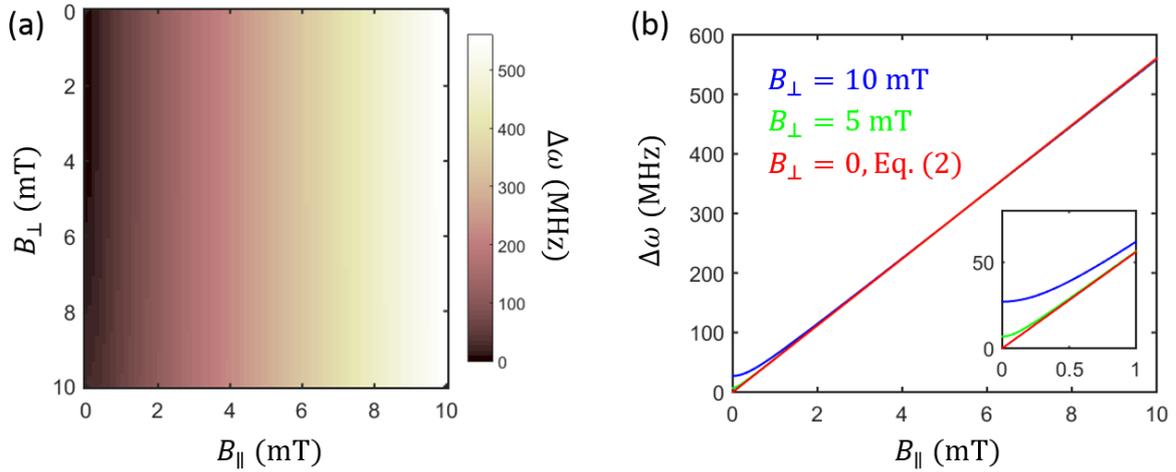

**Figure S3 – Zeeman splitting as a function of magnetic field.** (a) Difference between the two resonance frequencies of the NV centre, $\Delta\omega = \omega_+ - \omega_-$, as a function of the parallel and perpendicular components of the magnetic field, $B_\parallel$ and $B_\perp$, with respect to the NV axis, calculated using the Hamiltonian expressed in Eq. (1). (b) $\Delta\omega$ as a function of $B_\parallel$ for $B_\perp = 0, 5$ and $10$ mT. The case $B_\perp = 0$ (red line) is formally identical to the approximation expressed by Eq. (2), i.e. $\Delta\omega \approx 2\gamma_e B_\parallel$. The deviation from the $B_\perp > 0$ lines is visible mainly for $B_\parallel < 2$ mT.

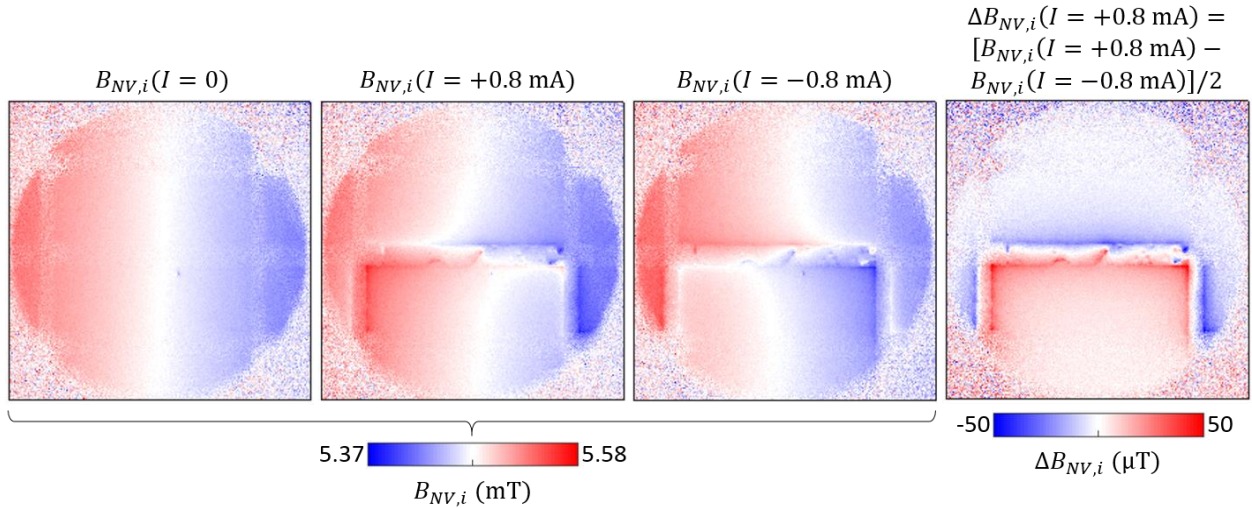

**Figure S4 – Subtraction of the background magnetic field.** Images of the magnetic field projection $B_{\text{NV},i=3}$ along the 3$^{\text{rd}}$ NV axis (as defined in Fig. S5), under different applied currents, $I$. The leftmost panel shows the case where no current is applied ($I = 0$), revealing the background magnetic field. The rightmost panel shows the difference between images with two currents of opposite signs, $\Delta B_{\text{NV},i}$, to leave only the contribution of the Oersted field while suppressing the background contribution.

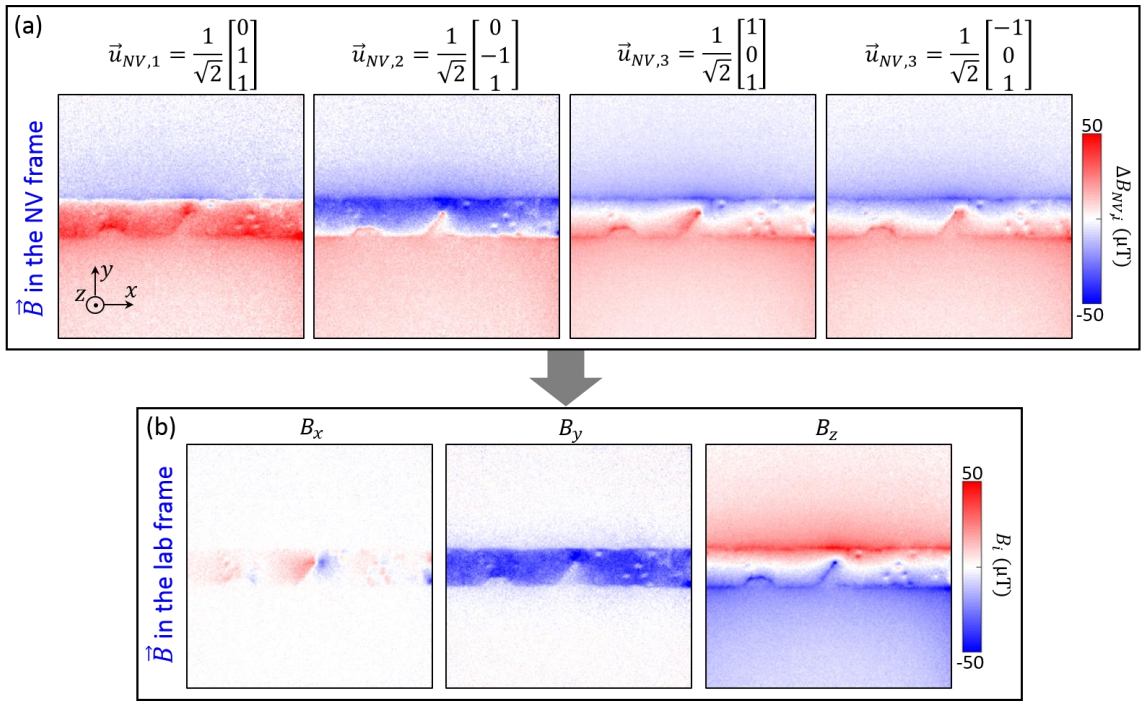

**Figure S5 – Oersted magnetic field in the lab frame.** (a) Projections of the Oersted magnetic field along the four NV axes, $\{\Delta B_{\mathrm{NV},i}\}$, under a current $I = 0.8$ mA. The Cartesian coordinates of the corresponding NV axis are indicated above each image, where the $xyz$ reference frame (the 'lab' frame) is defined in the leftmost panel. (b) Components of the Oersted magnetic field in the lab frame, $\{B_x, B_y, B_z\}$, obtained from (a).

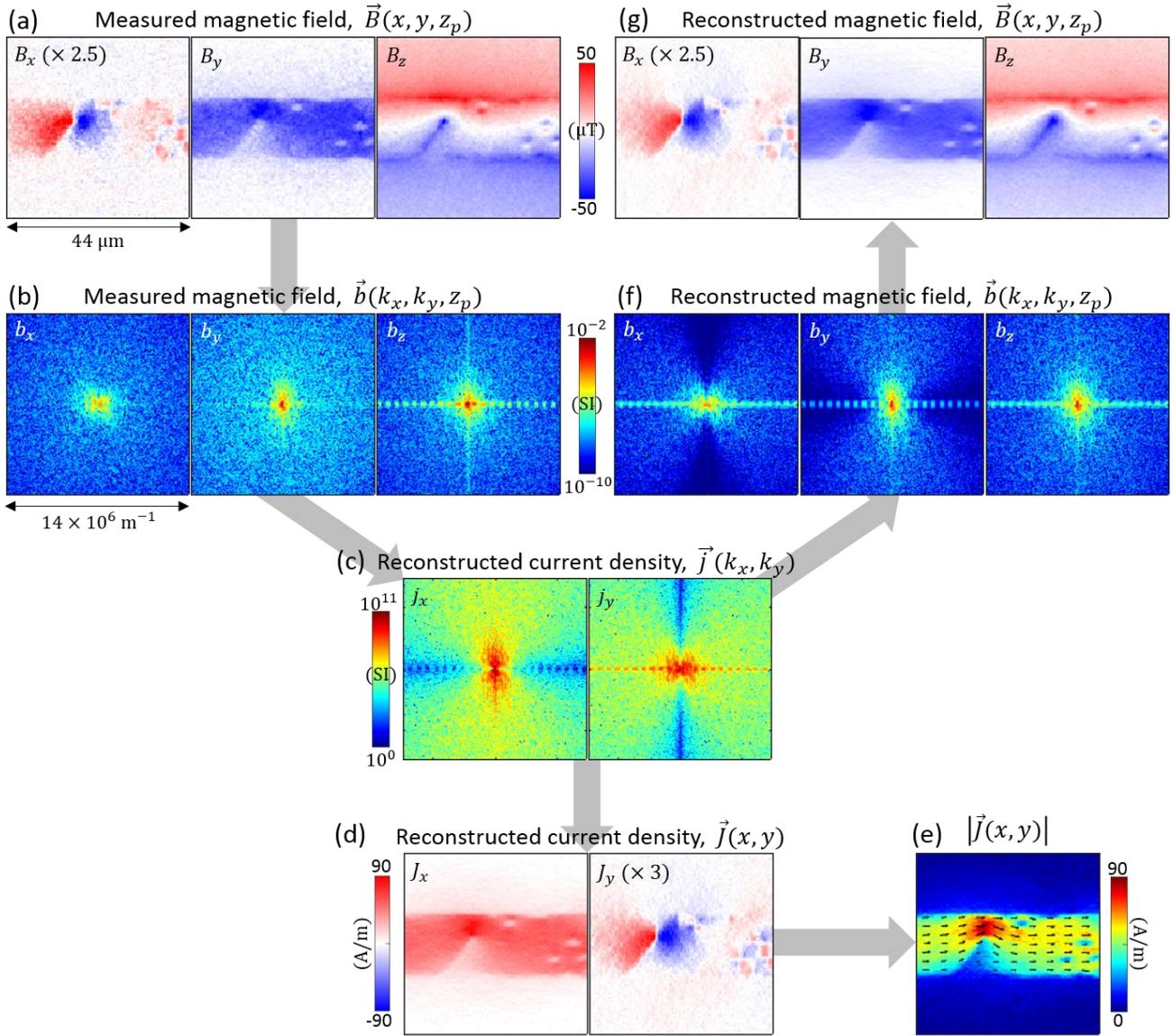

**Figure S6 – Procedure to reconstruct the current density.** Images of a graphene ribbon illustrating the reconstruction procedure. The injected current is $I = 0.8$ mA. The probe distance is taken to be $z_p = 20$ nm. (a) Measured magnetic field in the real space. (b) Fourier transforms of the images in (a). (c) Current density in the reciprocal space deduced from (b) as explained in the text. (d) Current density in the real space, obtained via inverse Fourier transform of (c). (e) Norm of the current density obtained from (d). The arrows represent the current density vector. (f) Magnetic field in the reciprocal space deduced from (c) using Eq. (11-13). (g) Magnetic field in the real space, obtained via inverse Fourier transform of (f). In (b,c,f), the centre of the images corresponds to $(k_x = 0, k_y = 0)$, and the images represent the absolute value of the Fourier transform, plotted in logarithmic scale.

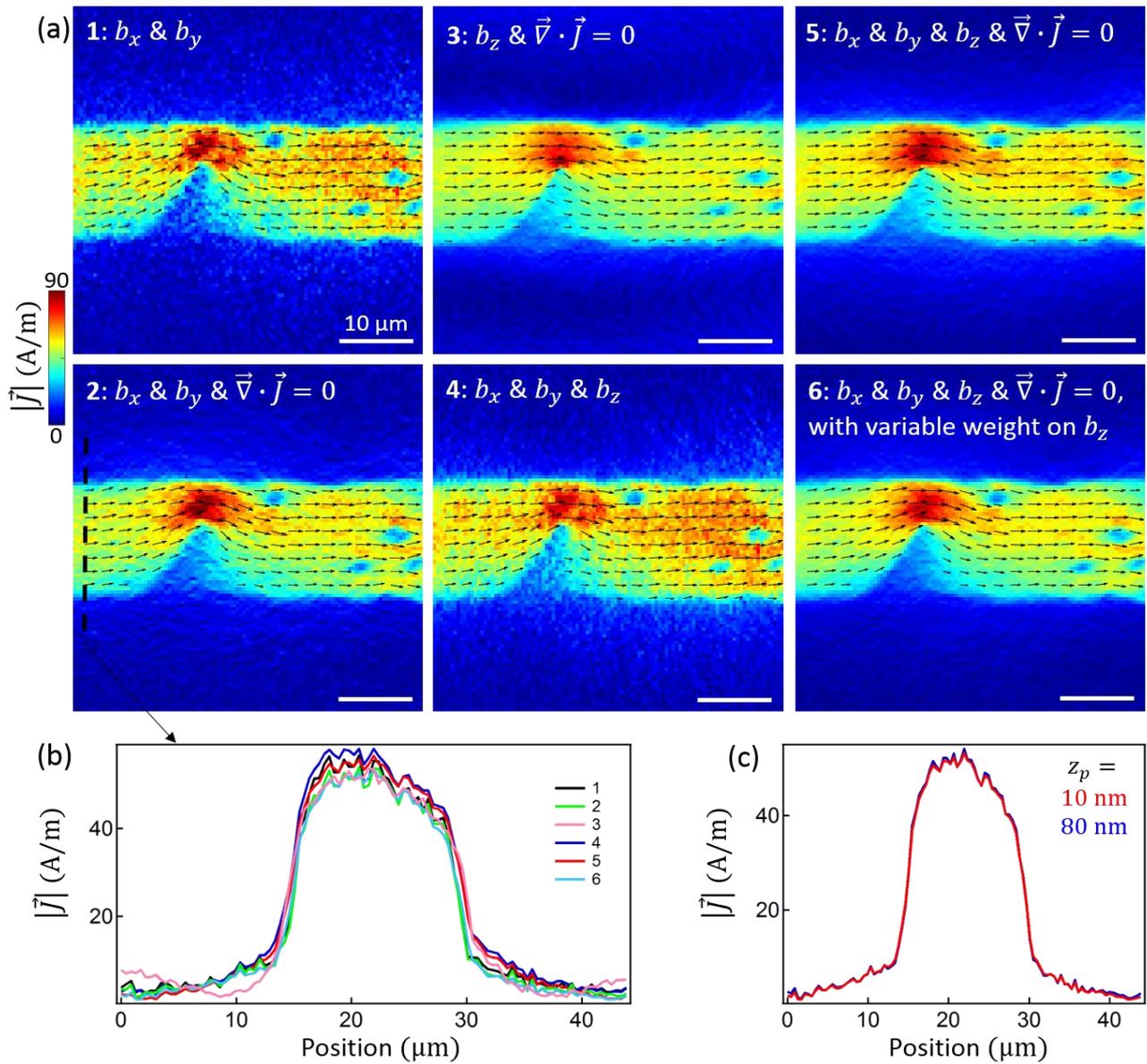

**Figure S7 – Robustness of the reconstruction procedure.** (a) Reconstructed density current using different sets of equations, as indicated on each image: $b_x$ refers to Eq. (11), $b_y$ refers to Eq. (12), $b_z$ refers to Eq. (13), and $\vec{\nabla} \cdot \vec{J} = 0$ is the continuity equation as expressed in Eq. (14). (b) Line cuts across the ribbon extracted from the images in (a). The labels 1 to 6 refer to those defined in (a). The line cuts are averaged over 10 pixels in width. (c) Line cuts obtained using the same sets of equations but assuming two different probe distances, $z_p = 10$ or $80$ nm.

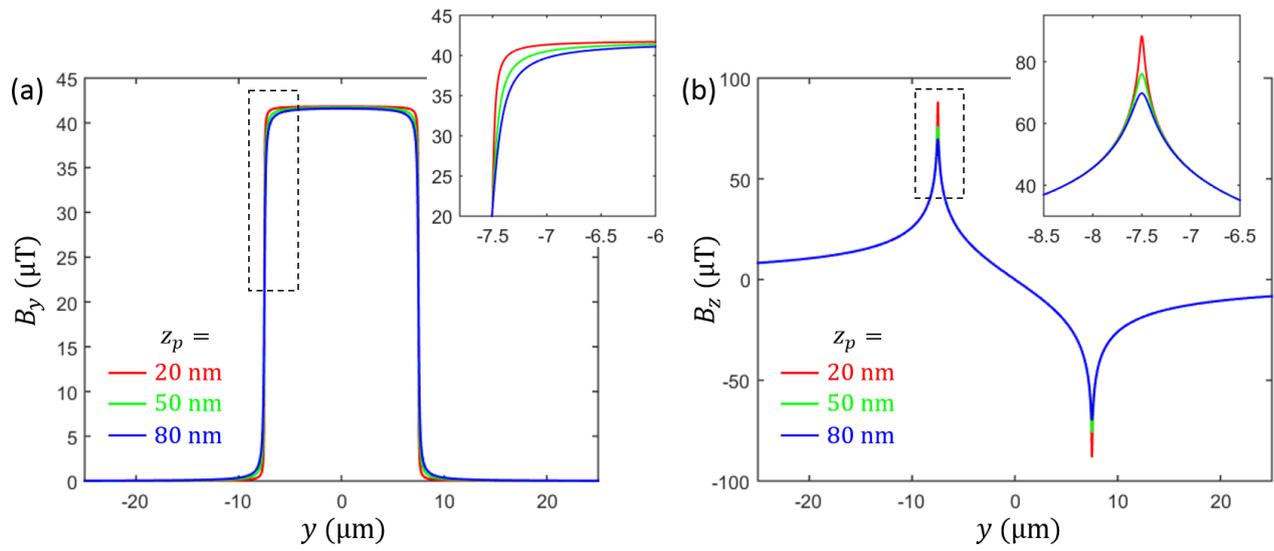

**Figure S8 – Oersted field as a function of probe distance.** (a,b) Components $B_y$ (a) and $B_z$ (b) of the Oersted magnetic field created by a current $I = 1$ mA in a ribbon of width $W = 15$ μm, computed as a function of the transverse position $y$ for various probe distances $z_p$ using Eqs. (16).